**Nanotechnology: a slightly different history**
Peter Schulz
School of Applied Sciences UNICAMP Brazil
peter.schulz@fca.unicamp.br

Many introductory articles and books about nanotechnology have been written to disseminate this apparently new technology, which investigate and manipulates matter at dimension of a billionth of a meter. However, these texts show in general a common feature: there is very little about the origins of this multidisciplinary field. If anything is mentioned at all, a few dates, facts and characters are reinforced, which under the scrutiny of a careful historical digging do not sustain as really founding landmarks of the field. Nevertheless, in spite of these flaws, such historical narratives bring up important elements to understand and contextualize this human endeavor, as well as the corresponding dissemination among the public: would nanotechnology be a cultural imperative?

### Introduction: plenty of room beyond the bottom

Nanotechnology is based on the investigation and manipulation of matter at the scale of billionths of a meter, i.e., nanometers; borrowing approaches from academic disciplines, which until recently were perceived more or less as isolated from each other: Biology, Physics, Chemistry and Material Sciences.

Different groups, ranging from the scientific community to the general public, when asked about the history of nanotechnology seem to be satisfied with a tiny set of information spread out from site to site in the web. Within the most disseminated allegedly origins of nanotechnology, we can find the so called grandfather and launching act of nanotechnology, namely the famous physicist Richard Feynman (1918 - 1988) and his 1959 talk *There is plenty of room at the bottom*[1]. We will come back to this below.

Afterwards, the word nanotechnology has been coined in 1974 by the Japanese researcher Norio Taniguchi (1912 - 1999). The "paternity" of nanotechnology is attributed to the first PhD recipient in the field, the American engineer Eric Drexler, author of *Engines of creation: the coming era of nanotechnology,* released in 1986, which became an important diffusion vehicle of this new technology to the public.

During the 1980´s, the fundamental discovery of molecules with 60 carbon atoms, the so called fullerenes, and the invention of scanning probe microscopes, turning possible the actual "atom by atom manipulation", would "open the doors to a new era".

Besides "modern" nanotechnology, there is also an "ancient" nanotechnology, going back to gold and silver nanoparticles responsible for special properties of some glasses produced during the Roman Empire. Evidently, those Romans had no idea that it was all about colloidal particles, now renamed as nanoparticles, systematically studied by Michael Faraday (1791 - 1867) already in the mid of the 19$^{th}$ century and sometimes mentioned in nanotechnology timelines.

Telling the story this way do not contribute to the understanding of nanotechnology as a human activity devoted to research and development with important social implications. Nevertheless, this tiny set of foot notes provides nice starting points for closer looks into the subject.

Let us recall the famous Feynman´s (Nobel Laureate in Physics, 1965) talk, asleep for over 20 years and switched into prophecy by, among others, Eric Drexler. Indeed, nothing better than a renowned oracle to foster an apparently new proposal.

The purpose of the talk was announced at its very beginning: "What I want to talk about is the problem of manipulating and controlling things on a small scale." Reading it further we get to know that small scale goes down to atomic scale. Just after this statement, a goal is also proposed: "Why cannot we write the entire 24 volumes of the *Encyclopedia Brittanica* on the head of a pin?" To answer this question Feynman brought together a series of conceptual possibilities which sound backwards as prophetic and, at that time were indeed very interesting.

But can we say these ideas were really visionary?

Looking to the context of the period, readers can reach their own conclusions, recalling that Feynman did not mention or cite scientific results in his talk, but he was probably a well-informed person. What was going on at that time? In 1958, a proof of concept of an integrated circuit, IC, was successfully presented and soon recognized as the first efficient route towards an unprecedented miniaturization of electronics. The American physicist and engineer Jack Kilby (1923 - 2005) - Nobel Laureate in Physics, 2000, for the invention of ICs – wrote in his lab notebook, in 1958: "Extreme miniaturization of many electrical circuits could be achieved by making resistors, capacitors, transistors and diodes on a single slice of Silicon".[2]

The word "extreme" opened a door for the imagination at this time in which competitions of miniaturization were hype, even before the prize offered by Feynman in his talk for the smallest motor of the world. So, Feynman had actually clear hints for his prophecy.

It is worth mentioning that Feyman´s talk has not influenced directly de development of nanotechnology in the way this field of knowledge actually evolved, as pointed out by the American Cultural Anthropologist Chris Toumey[3]. The real motivations behind Feynman´s, revisited also by Physics historians, do not fit into the a posteriori narrative constructed by the last decades of the last century[4]. Even so, Feynman´s talk reached superlative late fame, summing thousands of citations. Anyway, deepening into this debate reveals many other forerunners[5] and in what follows some of them are recalled.

On the other hand: would it be possible to an article barely known nowadays have had in fact direct influence on the development of nanotechnology? A possible yes as an answer to this question would be the paper *Molecular engineering* by Arthur von Hippel[6] (1898 – 2003), published in 1956, hence three years prior to Feynman´s talk.

**Molecular engineering**

Hippel´s manifesto like article, after a brief introduction, poses a question: "What is molecular engineering?" The answer is the very definition of nanotechnology: "…Instead of taking prefabricated materials and trying to devise engineering applications consistent with their macroscopic properties, one builds materials from their atoms and molecules for the purpose at hand…[the engineer] can play chess with elementary particles according to prescribed rules until new engineering solutions become apparent."

The conceptual developments suggested in the article in order to achieve such bold objective are rather modest comparing to Feynman´s. However, von Hippel addresses the question of which institutional framework is necessary to tackle this new king of engineering: "What we try to create as our answer to this situation are truly interdepartmental laboratories for molecular science and engineering." At that time, von Hippel headed a laboratory at MIT, with a "staff that consists of physicists, chemists, electrical engineers, and ceramists; we hope to form an alliance with mechanical and chemical engineers, metallurgists, and biologists as experience and confidence grow." Hence, von Hippel not only defined the scope of nanotechnology, but also anticipated the associated research environment, markedly interdisciplinary as observed today.

An interesting history of molecular engineering from von Hippel´s days, including the financing of ambitious projects aiming the substitution of Silicon even before the invention of the microprocessors, to the present is offered by Hyungsub Choy and Cyrus Mody in *"The long history of molecular electronics: Microelectronics Origins of nanotechnology"*.[7]

**Colloids Science**

What is the role of science diffusion books in promoting a new knowledge field? Drexler´s book, mentioned above, reminds us another one, now over 100 years old: *The world of neglected dimensions*, written in 1914 by the chemist Wolfgang Ostwald (1883 - 1943). With this book we come back to colloidal particles with dimensions ranging from micron down to nanometers and were not a mere curiosity at the beginning of the last century.

In order to promote this field of knowledge[8], Ostwald said that he "did not know a field of contemporary science which deals with so many and distinct fields of interest like chemistry of colloids. Surely that at that time radioactivity and atomic theory would get the attention from informed public, but are intellectual spices compared to the chemistry of colloids, which is necessary to many theoretical and practical areas."

This science of colloids reached its climax of academic perception with a sequence of Nobel prizes devoted to this field. First, the prize for Chemistry in 1925, to the Austrian chemist Richard Zsigmondy (1865 - 1928) and followed in the next year by the prize to the Swedish chemist Theodor Svedberg (1884 – 1971) and to the French physicist Jean Perrin (1870 – 1942).

Furthermore, there was a whole interdisciplinary research project, looking for technological applications, based on nanoparticles as pointed out by Gerald Holton: "It was widely thought that research on the colloidal state (the dispersed state of matter where particle dimensions are between $10^{-4}$ and $10^{-7}$ cm) was a great frontier for both pure and applied science, one that might bridge organic and inorganic matter. This field seemed filled with promise for medical-biological research as well as for industry".[9]

In other words, again we have a definition very close to the promises and fronts attributed to contemporary nanotechnology.

**Magic bullet**

Nanoparticles are also associated to systems of drug delivery: medicine assembled in nanoparticles embedded by a functionalized material which promises selectivity in looking for a specific target, like an ill cell, being more efficient, lower doses and less collateral effects[10]. Announced frequently as a revolution made possible by nanotechnology, but not yet satisfactorily achieved, this idea can be traced back to the beginning of the twentieth century, when the concept of "magic bullet"[11] was proposed by the German scientist Paul Erhlich ( 1854 – 1915), Nobel laureate for medicine in 1908: drugs that go only and directly to ill cells.

The development of strategies for obtaining such "magic bullets" where continuously searched since then: a very illustrative case is the use of gold and radioactive colloidal gold recovered with silver for mitigation of ascites and pleural effusions"[12], published in 1958. What is this work about? Radioactive gold has a therapeutic effect on diseases announced in the article and the authors found out that these nanoparticles would only reach the effected region if covered with silver. In summary: an example of "magic bullet" approach made possible by nanotechnology of the 1950´s and even commercialized at that time.

This is actually only one of the examples of early nanotechnology involving silver in nanoparticles. There are many others and nanoparticulate silver shows a systematic development over 120 years[13], with a broad agenda with many aspects of modern concerns, including toxicology, going back to the 1930s, which had to be rediscovered.

**Cultural imperative**

What those examples described above could have in common? We saw that the "contemporary agenda" of nanotechnology has been proposed much earlier at least twice, anticipating the necessity of interdisciplinary institutional frameworks, as well as ambitious founding projects.

Richard Jones[14] argues that "rather than considering this as the emergence of a new scientific field, nanotechnology is best thought of as a socio-political project that has arisen as a result of influences both from within science, and from the wider political, economic and cultural climate". This is very close to what could be said from the molecular engineering and colloidal science examples presented above.

Therefore, one could wonder if we are not witnessing just a new wave of nanotechnology without giving the proper credit to the previous ones. Hence, maybe we should consider nanotechnology as a whole, beginning actually with the chemistry of colloids and the "magic bullet" concept.

In this perspective, nanotechnology would be a "cultural imperative", concept introduced in cultural archeology by Michael Schiffer[15], who illustrated the concept by applying it to the pocket radio, a remarkable example of miniaturization with some similarities with the present discussion, in spite of the obvious difference in length scales.

Cultural imperatives are "mandates for a technological development", "a product fervently believed by a group – its constituency – to be desirable and inevitable, merely awaiting technological means for its realization". In such groups, promoters of the idea assume an outstanding role (Ostwald, von Hippel, Feynman, Drexler), as well as laymen and media, important for the diffusion of the idea, like the magazines about radio and electronics of the beginning of the twentieth century, in the case analyzed by Schiffer, who contribute to maintain the interest, besides the proper researchers and technical staff.

A cultural imperative may take decades to become viable and could go through different independent routes, like happens with nanotechnology. If a route becomes disseminated, there is a tendency to minimize the role of the previous ones, taking advantage of images that may not go beyond proofs of concept, but guarantee the public interest in, for instance, the current "third wave of nanotechnology".